\documentstyle[emulateapj,psfig]{article}

\makeatletter

\newenvironment{inlinefigure}{%
\def\@captype{figure}%
\noindent\begin{minipage}{0.999\linewidth}\begin{center}}
{\end{center}\end{minipage}\smallskip}
\makeatother

\lefthead{Steffen et al.}

\received{2003 June 12}
\begin{document}
\title{The Changing AGN Population}
\author{A.~T.~Steffen,$\!$\altaffilmark{1}
A.~J.~Barger,$\!$\altaffilmark{1,2,3}
L.~L.~Cowie,$\!$\altaffilmark{3}
R.~F.~Mushotzky,$\!$\altaffilmark{4}
Y.~Yang$\!$\altaffilmark{4,5}
}

\altaffiltext{1}{Department of Astronomy, University of Wisconsin-Madison,
475 North Charter Street, Madison, WI 53706.}
\altaffiltext{2}{Department of Physics and Astronomy, University
of Hawaii, 2505 Correa Road, Honolulu, HI 96822.}
\altaffiltext{3}{Institute for Astronomy, University of Hawaii,
2680 Woodlawn Drive, Honolulu, Hawaii 96822.}
\altaffiltext{4}{Laboratory for High Energy Astrophysics,
Goddard Space Flight Center, Code 660, NASA, Greenbelt, MD, 20770.}
\altaffiltext{5}{Department of Astronomy, University of Maryland,
College Park, MD, 20742.}

\slugcomment{Accepted by The Astrophysical Journal Letters}

\begin{abstract}
We investigate how the fraction of broad-line sources in the
AGN population changes with X-ray luminosity and redshift. We first 
construct the rest-frame hard-energy ($2-8$~keV) X-ray luminosity 
function (HXLF) at $z=0.1-1$ using {\it Chandra} 
Lockman Hole-Northwest wide-area data, {\it Chandra} Deep Field-North 
2~Ms data, other {\it Chandra} deep field data, and the {\it ASCA} 
Large Sky Survey data. We find that broad-line AGNs dominate above 
$\sim 3\times10^{43}$ ergs s$^{-1}$ and have a mean luminosity 
of $1.3\times10^{44}$ ergs s$^{-1}$. Type II AGNs can 
only become an important component of the X-ray population at
Seyfert-like X-ray luminosities. We then construct $z=0.1-0.5$ 
and $z=0.5-1$ HXLFs and compare them with both the 
local HXLF measured from {\it HEAO-1} A2 survey data and 
the $z=1.5-3$ HXLF measured from soft-energy ($0.5-2$~keV)
{\it Chandra} and {\it ROSAT} data. We find that the number 
density of $L_x>10^{44}$~ergs~s$^{-1}$ sources (quasars) steadily 
declines with decreasing redshift, while the number density of
$L_x=10^{43}-10^{44}$~ergs~s$^{-1}$ sources peaks at $z=0.5-1$.
Strikingly, however, the number density of broad-line AGNs 
remains roughly constant with redshift while their average luminosities
decline at the lower redshifts, showing another example of cosmic 
downsizing.
\end{abstract}


\section{Introduction}
\label{secintro}

It is becoming increasingly clear that low-redshift accretion
may be the dominant phase of supermassive black hole (SMBH) 
growth (\markcite{barger01}Barger et al.\ 2001;
\markcite{cowie03}Cowie et al.\ 2003). Although theoretical 
models of SMBH formation have relied on comparisons 
with the optical quasar luminosity function, optically
accessible broad-line active galactic nuclei (AGNs) form
only about 30\% (\markcite{barger03b}Barger et al.\ 2003b) 
of the X-ray background (XRB). 
Explanations for the lack of broad lines in most AGN optical
spectra include absorption along the line-of-sight or dilution of 
the emission-line signatures by stellar light
(\markcite{moran02}Moran, Filippenko, \& Chornock 2002).
Regardless of the reason, optically and soft X-ray 
selected samples substantially undercount the AGN population 
relative to hard X-ray selected samples, at least at intermediate 
luminosities (\markcite{cowie03}Cowie et al.\ 2003;
\markcite{barger03a}Barger et al.\ 2003a).

The hardest band for which the hard X-ray luminosity function (HXLF)
can presently be determined is rest-frame $2-8$~keV.
The $\approx2$~Ms exposure of the {\it Chandra} Deep Field-North 
(CDF-N) samples a large, distant cosmological volume down to a 
very faint hard X-ray flux limit ($f_{2-8~{\rm keV}} \approx1.4\times 
10^{-16}$~ergs~cm$^{-2}$~s$^{-1}$; 
\markcite{alexander03}Alexander et al.\ 2003). However,
this is substantially fainter than the X-ray flux of the
sources that contribute the most to the XRB 
(\markcite{cowie02}Cowie et al.\ 2002), and the low-redshift
volume sampled is small. To sample a large cosmological 
volume at low redshifts, we have mapped with {\it Chandra} an
$\sim0.4$~deg$^2$ area in the Lockman Hole-Northwest (LH-NW) 
field to a flux level of
$f_{2-8~{\rm keV}} \approx3\times 10^{-15}$~ergs~cm$^{-2}$~s$^{-1}$
(\markcite{yang03}Yang et al.\ 2003),
where most of the hard XRB is resolved. 
In this paper we use our LH-NW data, together with deep 
{\it Chandra} data on the CDF-N, A370, SSA13, and SSA22 fields 
and bright {\it ASCA} data from 
\markcite{akiyama00}Akiyama et al.\ (2000), to construct
rest-frame $2-8$~keV luminosity functions at low redshifts.
We then investigate the evolution with redshift of AGN populations 
separated by optical spectral type. 
We assume $\Omega_M=1/3$, $\Omega_\Lambda=2/3$, and 
$H_o=65$~km~s$^{-1}$~Mpc$^{-1}$. 
We use $L_x$ to denote the rest-frame $2-8$~keV luminosity.

\section{X-ray Sample Selection}
\label{secsamp}

The LH-NW data are taken from the X-ray and optical catalogs
of Y.~Yang et al.\ (in preparation) and
A.~T.~Steffen et al.\ (in preparation).
Here we consider only sources with fluxes 
above $5\times 10^{-16}$~ergs~cm$^{-2}$~s$^{-1}$ ($0.5-2$~keV) and
$5\times10^{-15}$ ($2-8$~keV) to provide good, high-significance 
samples. We also only consider those sources that have been targeted
spectroscopically. The only source property we considered when
making our masks for spectroscopy was X-ray flux (sources with higher 
X-ray fluxes had higher priority when we had to choose one X-ray 
source over another due to mask conflicts). 
We spectroscopically observed 150 sources out of our total
$2-8$~keV sample of 228. We were able to confidently
identify 100 of these 150 sources (67\%) based on
multiple emission and/or absorption line features.
Two of the identified sources are stars.  A practical optical
limit for obtaining redshifts for X-ray sources using the Deep 
Extragalactic Imaging Multi-Object Spectrograph (DEIMOS; 
\markcite{faber02}Faber et al.\ 2002) on Keck~II is about $R=24$
(see \markcite{barger03b}Barger et al.\ 2003b). Other
spectra may not have been identifiable due to slits falling too 
close to an edge of the CCD or bright neighbor objects 
compromising the spectra. 

%
%
\begin{figure*}[th]
\centerline{\psfig{figure=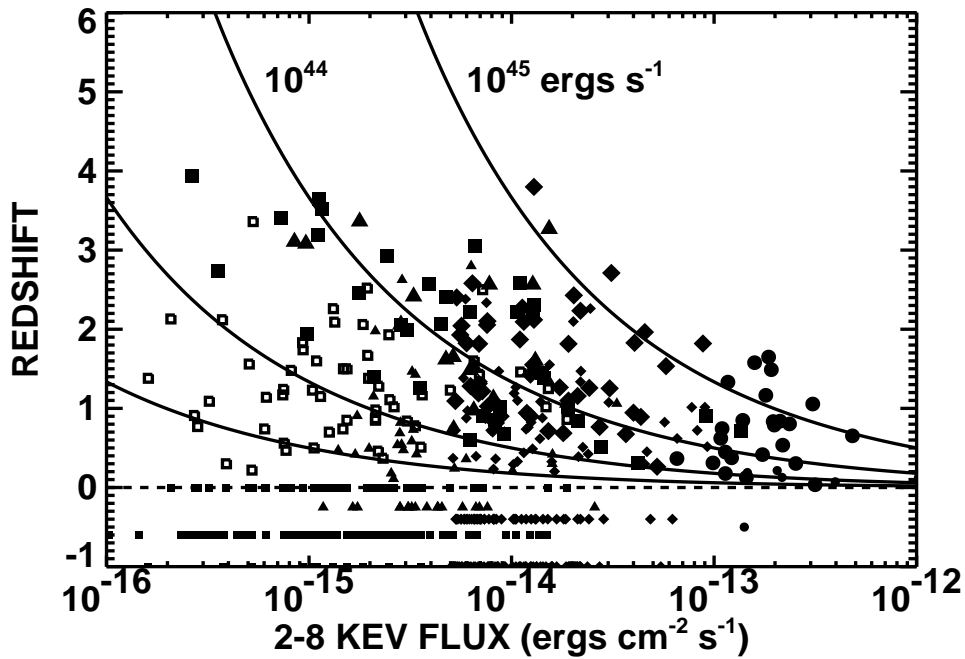,angle=90,width=5.3in}}
\vspace{6pt}
\figurenum{1}
\caption{
Redshift (solid for spectroscopic, open for photometric)
vs. $2-8$~keV flux for the X-ray samples
({\it squares}---CDF-N;
{\it triangles}---A370, SSA13, and SSA22;
{\it diamonds}---LH-NW; {\it circles}---{\it ASCA}).
Broad-line optical spectra are denoted by large solid
symbols. Unidentified sources are denoted by small solid
symbols at $z<0$. Solid curves show flux vs. redshift
for $L_x=10^{42}$~ergs~s$^{-1}$ (lowest curve),
$10^{43}$, $10^{44}$, and $10^{45}$ (highest curve),
computed with a $K$-correction for a $\Gamma=1.8$
power law spectrum (see Barger et al.\ 2002).
\label{fig1}
}
\end{figure*}

The faintest X-ray data are taken from the $\approx 2$~Ms CDF-N 
catalogs of \markcite{alexander03}Alexander et al.\ (2003; X-ray)
and \markcite{barger03b}Barger et al.\ (2003b; optical). 
Excluding stars and the small number of extended X-ray 
sources, the $2-8$~keV sample contains 326 sources,
68\% of which have either spectroscopic (165) or reliable 
photometric (56) redshifts. Our other {\it Chandra} and {\it ASCA} 
samples are summarized in \markcite{cowie03}Cowie et al.\ 
(2003). Hereafter, we refer to the A370, SSA13, and SSA22 
{\it Chandra} samples collectively as ``other''.

In our analysis we place every X-ray source that
was not spectroscopically identified for any reason into each
and every redshift interval in order to obtain a strict upper
bound on the number of AGNs that could lie in that redshift 
interval, modulo the existence of a substantial population of 
Compton-thick sources not detected by {\it Chandra}. We note in 
passing that \markcite{barger02}Barger et al.\ (2002) found from
the 1~Ms CDF-N data that a substantial population of Compton-thick 
sources is not needed to explain the greater than $8$~keV XRB, 
provided that the {\it ASCA} and {\it BeppoSAX} XRB measurements, 
instead of the lower {\it HEAO-1} A2 measurement, are assumed 
at energies less than $8$~keV.
For the LH-NW field, we only ever consider the sources that 
were spectroscopically observed (identified or not). This 
is acceptable because the spectroscopy are purely from targeted 
observations of the X-ray samples. For the other fields, the 
spectroscopy are not purely from targeted observations of 
the X-ray samples, so we need to include all of the sources.

We determined the solid angle covered by the combined samples 
at a given flux by comparing the observed numbers of sources
versus flux with the averaged number counts in the
$2-8$~keV band from \markcite{cowie02}Cowie et al.\ (2002).
This method allows a simple treatment of the incompleteness
that was modeled in computing the counts. However, the counts 
in Cowie et al.\ also include the low CDF-S counts,
which may affect the normalization at the 10\% level relative
to the LH-NW average (\markcite{yang03}Yang et al.\ 2003). We 
consider this to be a reasonable estimate of the systematic 
errors in the present analysis. The solid angle covered by the 
combined $2-8$~keV samples ranges from 0.002~deg$^2$ at
the faintest fluxes to 5.8~deg$^2$ at the highest fluxes.
At $2.3\times 10^{-14}$~ergs~cm$^{-2}$~s$^{-1}$ ($2-8$~keV) the 
solid angle is 0.52~deg$^2$.

In Figure~\ref{fig1} we show redshift versus $2-8$~keV flux
for the {\it ASCA} ({\it circles}), LH-NW ({\it diamonds}),
CDF-N ({\it squares}), and other ({\it triangles}) samples.
The LH-NW sources nicely fill in the almost an order
of magnitude flux gap between the CDF-N and {\it ASCA} samples.
The solid curves correspond to loci of constant $L_x$.
Any source more luminous than $L_x=10^{42}$~ergs~s$^{-1}$ is
very likely to be an AGN on energetic grounds
(\markcite{zezas98}Zezas, Georgantopoulos, \& Ward 1998;
\markcite{moran99}Moran et al.\ 1999),
though many of the intermediate luminosity sources
do not show obvious AGN signatures in their optical spectra.

\section{The AGN X-ray Luminosity Function}
\label{seclf}

\markcite{cowie03}Cowie et al.\ (2003) used {\it Chandra},
{\it ASCA}, and {\it ROSAT} data to determine rest-frame
$2-8$~keV AGN luminosity functions in two redshift intervals,
$z=0.1-1$ and $z=2-4$. They did not have very many sources in 
their sample with $2-8$~keV fluxes $\sim 2\times 10^{-14}$ to
$10^{-13}$~ergs~cm$^{-2}$~s$^{-1}$ since that
intermediate flux range is not well covered by either small-area,
deep {\it Chandra} observations or bright {\it ASCA} observations.
With our wide-area LH-NW data, we now have good coverage of the
intermediate flux range and can therefore improve the determination 
of the $z=0.1-1$ HXLF.

%
%
\begin{inlinefigure}
\psfig{figure=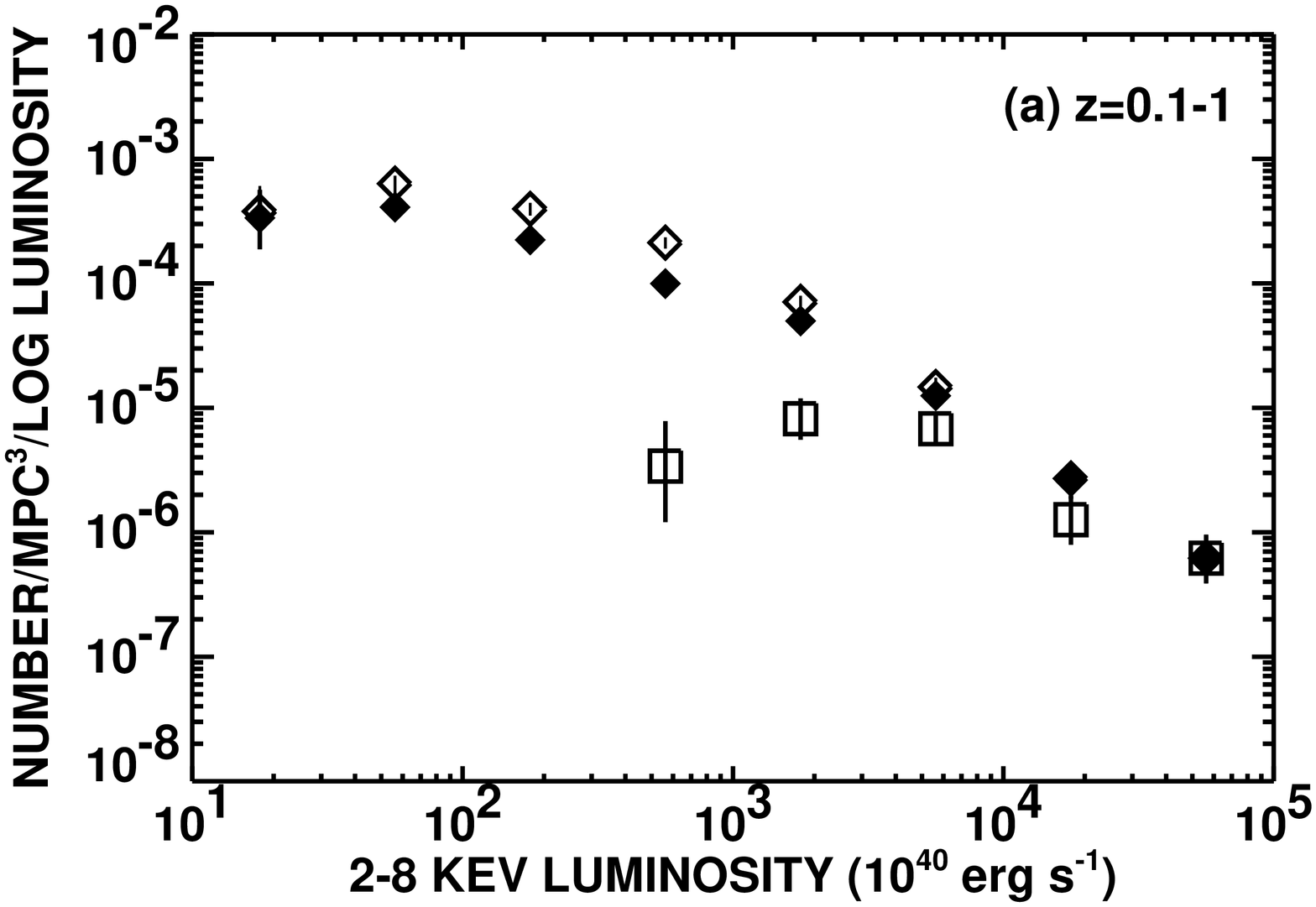,angle=90,width=3.5in}
\psfig{figure=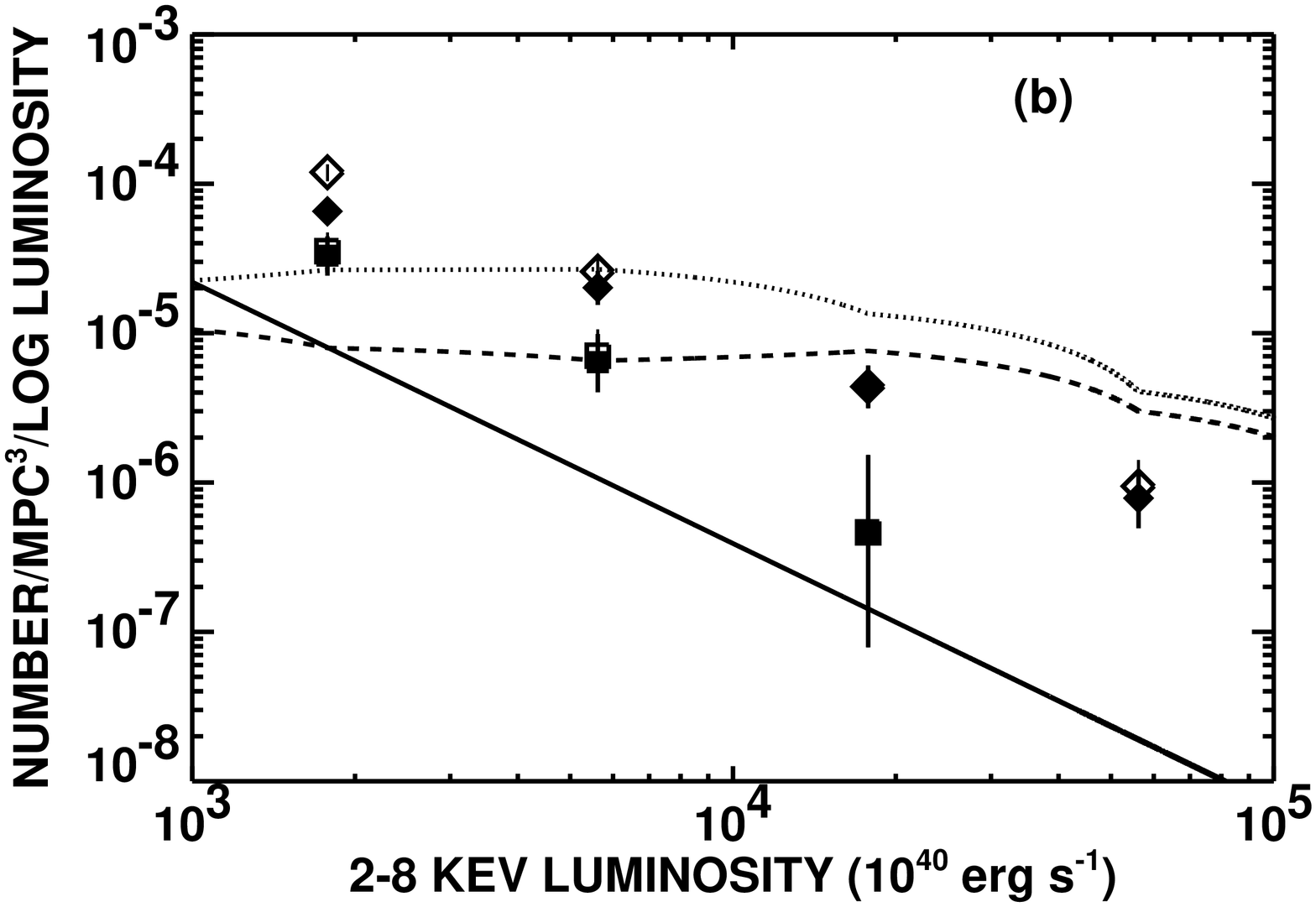,angle=90,width=3.5in}
\vspace{6pt}
\figurenum{2}
\caption{
(a) Rest-frame $2-8$~keV luminosity function per unit logarithmic
luminosity for $z=0.1-1$ ({\it solid diamonds}), computed from
observed-frame $2-8$~keV assuming an intrinsic $\Gamma=1.8$,
for which there are only
small differential $K$-corrections to rest-frame $2-8$~keV.
No correction was made for instrinsic absorption, but
in rest-frame $2-8$~keV, average absorption corrections are small.
Maximal $z=0.1-1$ HXLF ({\it open diamonds}) was found by assigning
all of the unidentified sources redshifts at the volume-weighted
centers of each and every redshift interval.
True HXLF lies somewhere between the solid and open diamonds.
Open squares show the $z=0.1-1$ HXLF for broad-line AGNs only.
Poissonian $1\sigma$ uncertainties are based on the
number of sources in each logarithmic luminosity bin.
(b) Rest-frame HXLF from (a) broken into two smaller redshift
intervals, $z=0.1-0.5$ ({\it solid squares}) and $z=0.5-1$
({\it solid diamonds}).
Maximal HXLFs are shown with open symbols. The
$z=1.5-3$ HXLF ({\it dashed line}; computed from observed-frame
$0.5-2$~keV) and the local Seyfert galaxy HXLF ({\it solid line};
from Piccinotti et al.\ 1982) are shown for comparison. Dotted line
is maximal $z=1.5-3$ HXLF, found by assigning redshifts to all of
the unidentified sources at the volume-weighted centers of
each and every redshift interval. True $z=1.5-3$ HXLF lies
somewhere between the dashed and dotted lines.
\label{fig2}
}
\addtolength{\baselineskip}{10pt}
\end{inlinefigure}

In Figure~\ref{fig2}a we show our measured HXLF
for the $z=0.1-1$ interval ({\it solid diamonds}), computed following
\markcite{cowie03}Cowie et al.\ (2003), who used the traditional
$1/V_a$ method of \markcite{felten77}Felten (1977).
The uncertainties are Poissonian, based on the number of galaxies
in each logarithmic luminosity bin.
However, since incompleteness is a potentially larger source of
error, we also recomputed the HXLF ({\it open diamonds}) by assigning
all of the unidentified sources to the volume-weighted centers of
each and every redshift interval. Because the spectroscopic
identifications are
much more complete at the high X-ray fluxes, the associated
systematic uncertainties are larger at low $L_x$.
The HXLF of only the broad-line AGNs is shown
by the large open squares in Figure~\ref{fig2}a. In the high
luminosity range where the HXLF is steep, we can see that
the broad-line AGNs dominate the total HXLF.

The uncorrected $z=0.1-1$ HXLF can be well represented by a
conventional broken power law fit of the form
$6.6\times 10^{-6}~L_{44}^{-1.37}$ per Mpc$^3$ above
$2\times 10^{43}$~ergs~s$^{-1}$ and
$1.5\times 10^{-5}~L_{44}^{-0.87}$ below, where $L_{44}$ is the
luminosity in units of $10^{44}$~ergs~s$^{-1}$.
The integrated $z=0.1-1$ light density is 
$1.8\times 10^{39}$~ergs~s$^{-1}$~Mpc$^{-3}$.
Half of the total uncorrected light in the $z=0.1-1$ interval 
arises in sources more luminous than $2.1\times 10^{43}$~ergs~s$^{-1}$, 
and about one-third arises in broad-line AGNs.

In Figure~\ref{fig2}b we show our measured HXLF over the luminosity 
range $L_x=10^{43}$ to $10^{45}$~ergs~s$^{-1}$, where most of the 
sources have been spectroscopically identified, divided into two 
smaller redshift intervals, $z=0.1-0.5$ ({\it squares}) and 
$z=0.5-1$ ({\it diamonds}). For comparison, we show
the local Seyfert galaxy HXLF ({\it solid line})
from \markcite{picc82}Piccinotti et al.\ (1982; converted to
$H_0=65$~km~s$^{-1}$~Mpc$^{-1}$ and, using their assumed $\Gamma=1.7$,
to $2-8$~keV) and the high-redshift HXLF ({\it dashed line})
from \markcite{cowie03}Cowie et al.\ (2003;
updated to include our LH-NW data and the 2~Ms CDF-N data,
and to cover the interval $z=1.5-3$). We have computed the maximal 
high-redshift HXLF ({\it dotted line}) by assigning 
all of the unidentified sources to the volume-weighted centers of 
each and
every redshift interval. The high-redshift HXLF is much tighter than
in Cowie et al.\ (2003) because of our slightly lower redshift 
interval and our higher spectroscopic completeness.
From Figure~\ref{fig2}b we see that the number density of
$L_x>10^{44}$~ergs~s$^{-1}$ sources
steadily declines with decreasing redshift. However,
the number density of $L_x=10^{43}-10^{44}$~ergs~s$^{-1}$ sources
is largest in the $z=0.5-1$ interval and smaller at both higher 
and lower redshifts (see also Fig.~\ref{fig4}).

\section{X-ray Fluxes and Luminosities}
\label{seclum}

Prior to {\it Chandra} and {\it XMM-Newton},
various groups 
(e.g., \markcite{comastri95}Comastri et al.\ 1995;
\markcite{gilli01}Gilli, Salvati, \& Hasinger 2001)
using XRB population synthesis models based on 
unified AGN schemes predicted that a mixture of unobscured 
type I AGNs and intrinsically obscured type II AGNs were needed 
to produce the bulk of the XRB.
The obscured systems were expected to lie at high redshifts 
($z=2-3$), with the high luminosity end (type II quasars) 
contributing substantially to the XRB.
Contrary to the predictions, optical observations of the 
sources detected in deep {\it Chandra} and {\it XMM-Newton} 
hard X-ray surveys have found a redshift distribution
that peaks at $z<1$ (\markcite{barger03b}Barger et al.\ 2002, 2003b;
\markcite{hasinger03}Hasinger 2003).
Only a handful of type II quasars (here we define quasars
as having $L_x>10^{44}$~ergs~s$^{-1}$) have been detected, and these 
contribute only a small fraction of the XRB
(e.g., \markcite{barger03}Barger et al.\ 2003b).

In Figure~\ref{fig3} we plot for the redshift intervals 
$z=0.1-1$ and $z=1.5-3$ the fractional number of X-ray sources 
in each luminosity bin, by optical spectral type, versus rest-frame 
$2-8$~keV luminosity.
From Figure~\ref{fig3}a we see that the dominant population
at quasar luminosities, where our redshift identifications
are very complete, is broad-line AGNs. However,
at more Seyfert-like luminosities ($L_x<10^{44}$~ergs~s$^{-1}$),
the broad-line AGN fraction declines rapidly with decreasing luminosity. 
Figure~\ref{fig3}b shows a similar trend with luminosity.
These results suggest a luminosity dependence in optical
spectral type: type I AGNs dominate at quasar luminosities, while 
type II AGNs can only become an important component of the X-ray 
population at Seyfert-like luminosities.

%
%
\begin{inlinefigure}
\psfig{figure=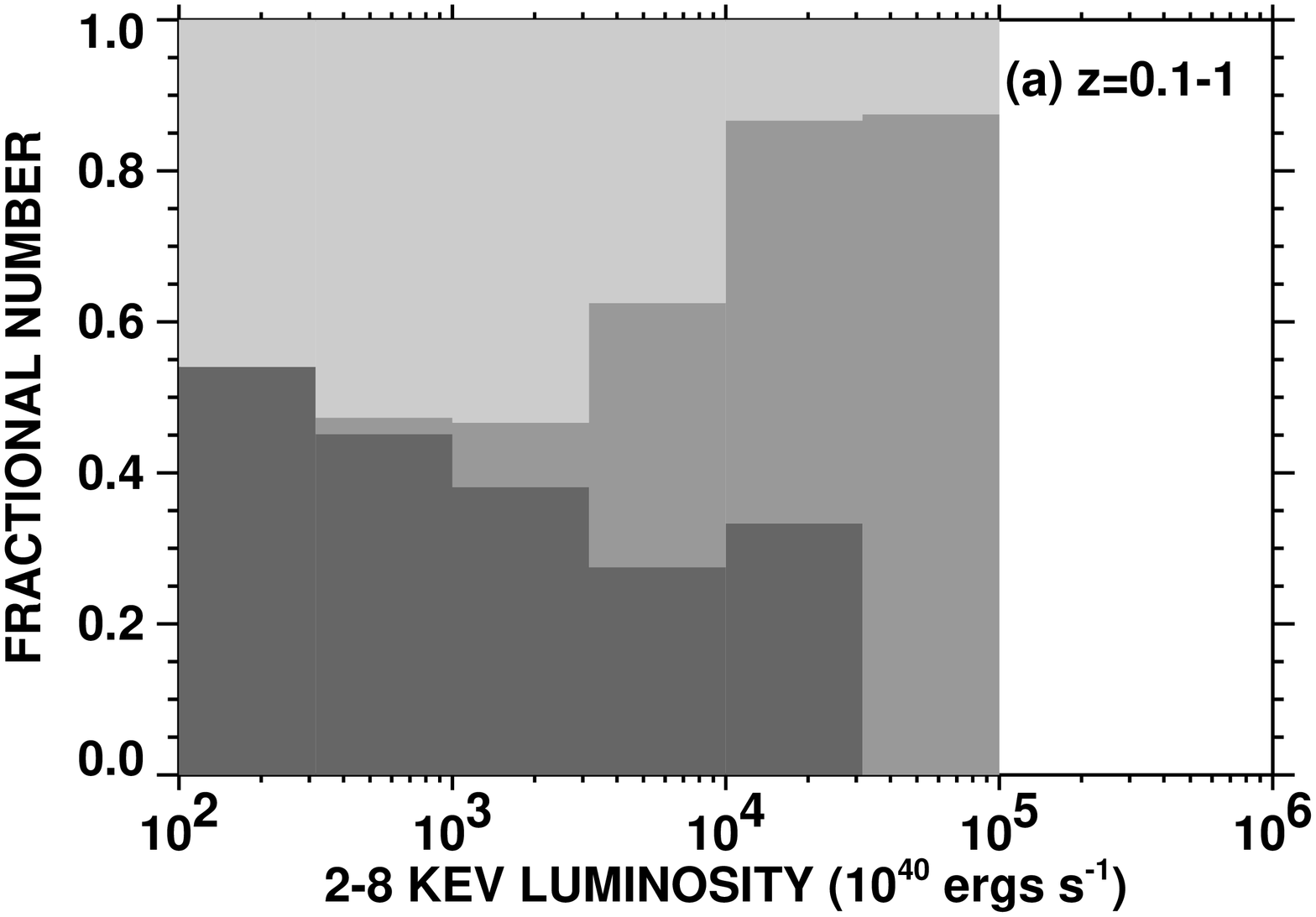,angle=90,width=3.5in}
\psfig{figure=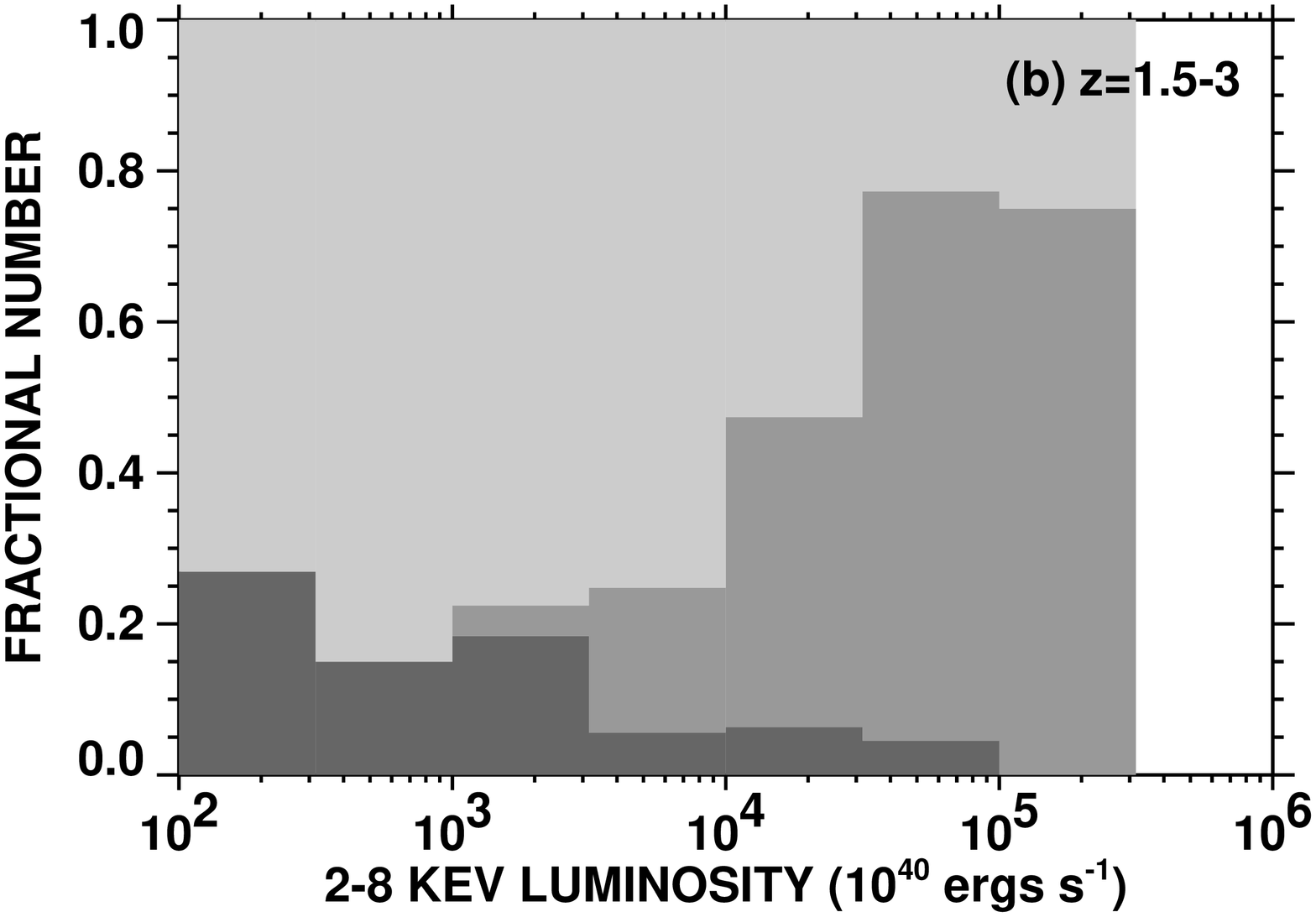,angle=90,width=3.5in}
\vspace{6pt}
\figurenum{3}
\caption{
Fractional number of X-ray sources by optical spectral type
({\it dark shading}---non-broad-line
sources; {\it medium shading}---broad-line AGNs;
{\it light shading}---unidentified sources)
versus $2-8$~keV luminosity for the redshift intervals
(a) $z=0.1-1$ and (b) $z=1.5-3$.
Luminosities were computed with a $K$-correction for a $\Gamma=1.8$
power law spectrum. All of the unidentified sources were assigned
redshifts at the volume-weighted centers of each and every redshift
interval.
\label{fig3}
}
\addtolength{\baselineskip}{10pt}
\end{inlinefigure}

To quantify this, we note that in the $z=0.1-1$ redshift interval 
there are 20 sources with quasar luminosities, of which 15 
are broad-line AGNs. Only three of the unidentified sources would 
have quasar luminosities if placed in this redshift interval, 
so the systematic errors are small, and the bulk of the quasars
are broad-line AGNs. In the $z=1.5-3$ interval
there are 65 sources with quasar luminosities, of which 58
are broad-line AGNs. There are 56 unidentified sources
that would have quasar luminosities if placed in this redshift
interval, so the systematic uncertainties are larger. However,
at least half (and probably considerably more) of the sources
with quasar luminosities in this redshift interval are
broad-line AGNs.

\section{Discussion}

\markcite{barger03a}Barger et al.\ (2003a) graphically showed
how the $L_x=10^{43}-10^{44}$~ergs~s$^{-1}$ 
(Seyfert-like luminosities) number density slowly increases 
with decreasing redshift from $z=6$ to $z=0.1$, while the 
$L_x=10^{44}-10^{45}$~ergs~s$^{-1}$ (quasar luminosities)
number density peaks at $z=1.5-3$. They also showed that the evolution 
of the $L_x=10^{44}-10^{45}$~ergs~s$^{-1}$ number density closely 
matches in shape the evolution of optically selected quasars.
Given these trends and our interesting result from Figure~\ref{fig3} 
that while the dominant population at quasar luminosities is 
broad-line AGNs, at lower luminosities the broad-line fraction drops 
dramatically, here we explore a different aspect of the number 
density evolution.

In Figure~\ref{fig4} we show number density versus redshift
for the sources with $L_x>10^{43}$~ergs~s$^{-1}$. Rather than
dividing the sources into luminosity classes, as was done in
\markcite{barger03a}Barger et al.\ (2003a), we have instead
divided the sources into broad-line ({\it open squares}) and
non-broad-line ({\it solid circles}) AGNs. The horizontal bars
show the maximal number densities for the non-broad-line AGNs,
obtained by assigning all of the unidentified sources redshifts at
the volume-weighted centers of each and every redshift bin and 
then retaining only those with $L_x>10^{43}$~ergs~s$^{-1}$.
The bars are not consistent with one another because
all of the unidentified sources are included in each and every 
redshift bin, as long as their luminosities at those redshifts 
exceed $10^{43}$~ergs~s$^{-1}$.

%
%
\begin{inlinefigure}
\psfig{figure=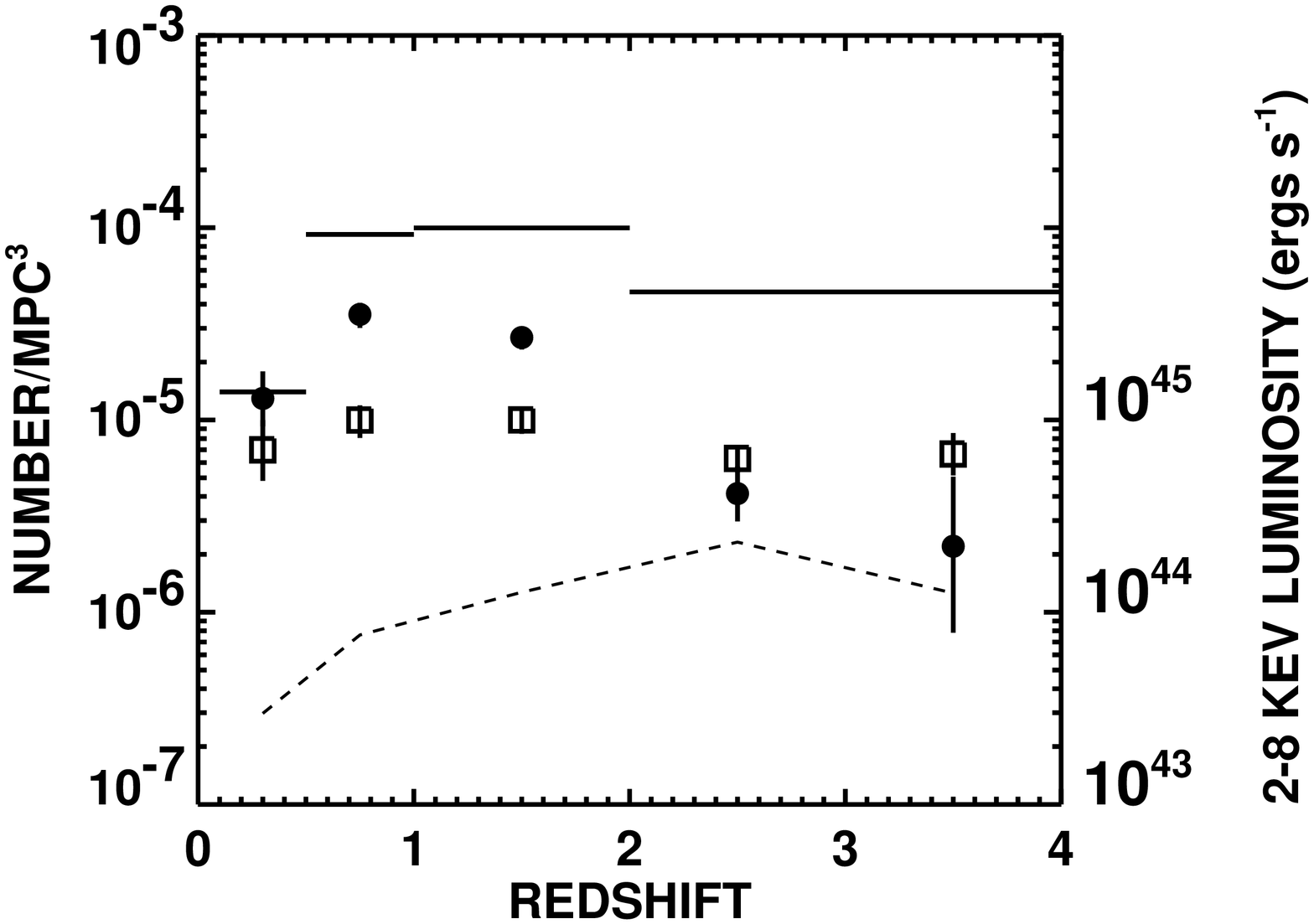,angle=90,width=3.5in}
\vspace{6pt}
\figurenum{4}
\caption{
Number density of sources with
$L_x>10^{43}$~ergs~s$^{-1}$ vs. redshift. Broad-line (non-broad-line)
AGNs are denoted by open (solid) symbols. Points above (below)
$z=2$ were determined from observed-frame $2-8$~keV
($0.5-2$~keV), assuming an intrinsic $\Gamma=1.8$.
Poissonian $1\sigma$ uncertainties
are based on the number of sources in each redshift bin.
Horizontal bars show the maximal number densities for the
non-broad-line AGNs, found by assigning redshifts to all of the
unidentified sources at the volume-weighted centers of each and
every redshift
bin and then retaining only those with $L_x>10^{43}$~ergs~s$^{-1}$.
Dashed line shows the mean luminosity (right-hand axis) of the
broad-line AGNs with redshift.
\label{fig4}
}
\addtolength{\baselineskip}{10pt}
\end{inlinefigure}

From Figure~\ref{fig4} we see that the number density of 
non-broad-line AGNs with $L_x>10^{43}$~ergs~s$^{-1}$ rises with 
decreasing redshift, until the last redshift bin. Strikingly, 
however, the number density of broad-line AGNs
remains roughly constant with redshift while their average 
luminosities ({\it dashed line; right-hand axis}) decline
at the lower redshifts. This is another example of cosmic downsizing. 
In the terminology of previous quasar studies, the broad-line AGNs 
are undergoing pure luminosity evolution, not density evolution. 
Pure luminosity evolution is consistent with the most recent
optical determinations (\markcite{boyle00}Boyle et al.\ 2000).
The increase in the number density of 
non-broad-line sources and the decrease in the broad-line AGN 
average luminosities results in non-broad-line sources 
increasingly dominating the light at lower redshifts, modulo
the existence of a substantial population of Compton-thick sources.

\acknowledgements
We gratefully acknowledge support from NASA's National Space
Grant College and Fellowship Program and the Wisconsin
Space Grant Consortium (A.~T.~S.), CXC grants
GO2-3191A (A.~J.~B.) and GO2-3187B (L.~L.~C.),
NSF grants AST-0084847 (A.~J.~B.) and AST-0084816 (L.~L.~C.),
the University of Wisconsin Research Committee with funds 
granted by the Wisconsin Alumni Research Foundation (A.~J.~B.), 
the Alfred P. Sloan foundation (A.~J.~B.), and 
the IDS program of R.~F.~M.

\end{document}